\documentclass[aps,prb,twocolumn,showpacs,superscriptaddress,groupedaddress]{revtex4}  
\usepackage{graphicx}  
\usepackage{dcolumn}   
\usepackage{bm}        
\usepackage{amssymb}   
\usepackage{framed}
\usepackage{amsmath}

\begin{document}

\widetext
\title{Benchmark tests of a strongly constrained semilocal functional with a long-range dispersion correction}

\affiliation{London Centre for Nanotechnology, 	, 17-19 Gordon Street, London WC1H 0AH, United Kingdom.}
\affiliation{Department of Physics, Temple University, Philadelphia, Pennsylvania 19122, USA}
\affiliation{Department of Chemistry, Temple University, Philadelphia, Pennsylvania 19122, USA}
\author{J.~G.~Brandenburg}
\email{g.brandenburg@ucl.ac.uk}
\affiliation{London Centre for Nanotechnology, University College London, 17-19 Gordon Street, London WC1H 0AH, United Kingdom.}
\author{J.~E.~Bates} \affiliation{Department of Physics, Temple University, Philadelphia, Pennsylvania 19122, USA}
\author{A.~Ruzsinszky} \affiliation{Department of Physics, Temple University, Philadelphia, Pennsylvania 19122, USA}
\author{J.~Sun} \affiliation{Department of Physics, Temple University, Philadelphia, Pennsylvania 19122, USA}
\author{J.~P.~Perdew} \affiliation{Department of Physics, Temple University, Philadelphia, Pennsylvania 19122, USA} \affiliation{Department of Chemistry, Temple University, Philadelphia, Pennsylvania 19122, USA}
                             
\date{\today}

\begin{abstract}
The strongly constrained and appropriately normed (SCAN) semilocal density functional
[J. Sun, A. Ruzsinszky, J. P. Perdew \textit{Phys. Rev. Lett.} {\bf 115}, 036402 (2015)]
obeys all 17 known exact constraints 
for meta-generalized-gradient approximations (meta-GGA) and includes some medium range correlation
effects. Long-range London dispersion interactions are still missing, but can be accounted for
via an appropriate correction scheme. 
In this study, we combine SCAN with an efficient London dispersion correction and show
that lattice energies of simple organic crystals can be improved with the applied correction by 50\%.
The London-dispersion corrected SCAN meta-GGA outperforms all other tested London-dispersion corrected meta-GGAs for molecular geometries.
Our new method delivers mean absolute deviations (MADs) for main group bond lengths that are consistently below 1\,pm,
rotational constants with MADs of 0.2\%, and noncovalent distances with MADs below 1\%.
For a large database of general main group thermochemistry and kinetics,
it also delivers a weighted mean absolute deviation
below 4 kcal/mol, one of the lowest for long-range corrected meta-GGA functionals.
We also discuss some consequences of numerical sensitivity encountered for meta-GGAs.
\end{abstract}

\maketitle

\section{\label{sec:intro}Introduction}
Kohn-Sham density functional theory (KS-DFT, or DFT in the following)\cite{dftbook,kohn98}
has become an irreplaceable tool for the calculation of electronic structure
in chemical and physical sciences.
Within KS-DFT a noninteracting system is introduced with an effective
one-particle Hamiltonian, $\hat{h}_{KS}$, whose ground state density $\rho$ is equivalent to the
interacting system.
The wavefunction of the auxiliary non-interacting system is an anti-symmetrized product 
of single-particle eigenfunctions $\psi_{i}$ (KS orbitals),
the solutions of a coupled set of non-linear equations
\small
\begin{align}
 \hat{h}_{KS}\psi_{i}(\mathbf{r}) =& \epsilon_i \psi_{i}(\mathbf{r})\\
 \hat{h}_{KS} =& \hat{T} + \hat{V}_{ext} + \hat{V}_{Coul} + \hat{V}_{xc}
\end{align}
\normalsize
with kinetic energy operator $\hat{T}$, external potential (typically describing the fixed nuclear charges) $\hat{V}_{ext}$,
the mean field Coulomb (or Hartree) potential $\hat{V}_{Coul}$, and the exchange-correlation (xc) potential $\hat{V}_{xc}$.

While DFT is in principle an exact theory, in practice the exchange-correlation energy has to be approximated.
Density functional approximations (DFAs) are constructed by satisfying known exact constraints, or by
empirical fitting.
There are three main classes of DFAs that use only the local density and other semi-locally-available information to approximate the xc energy, $E_{xc}$.
The first is the local spin density approximation (LSDA), which is exact for the uniform electron gas.\cite{vbh72}
LSDA is still widely used in the solid state community with recent extensions to finite temperature free energies.\cite{trickney-lsda}
While extended metallic systems can be described reasonably well by LSDA, 
typical molecular systems require inclusion of the density gradient as in
the generalized gradient approximation (GGA). 
The most prominent GGAs are the Perdew-Burke-Enzerhof (PBE) exchange and correlation functionals\cite{pbe}
and the Becke exchange (B88)\cite{b88} combined with the Lee-Yang-Parr (LYP) correlation functional.\cite{lyp}
A natural extension to GGAs is to use higher-order derivatives of the electron density or
other semilocally-available information, leading to the meta-GGA class.
A typically employed variable is the KS kinetic energy density $\tau=\frac{1}{2}\sum_i |\nabla \psi_i|^2$.
Popular meta-GGAs are the Tao-Perdew-Staroverov-Scuseria (TPSS) functional\cite{tpss} 
and the Minnesota functionals M06L,\cite{m06l} M11L,\cite{m11l} and MN12L\cite{mn12l} by Truhlar and coworkers.
A recently introduced
empirical meta-GGA with a smoothness constraint
and a VV10 long-range dispersion correction, B97M-V, was presented by Mardirossian and Head-Gordon.\cite{b97mv}
Constraint-satisfaction based meta-GGA functionals have gained more attention in recent years\cite{trickney-metagga,ms0}. 
The here-analyzed SCAN functional is also a meta-GGA.\cite{scan}

In contrast to the empirical design of the Minnesota functionals, SCAN 
was built to satisfy the 17 known exact constraints for a semilocal functional
using appropriate norms for different limits.
SCAN has been shown to be superior to PBE for some standard molecular and solid test sets.\cite{scan,natchem} 
It is the first efficient functional 
that demonstrates simultaneous accuracy for 
diversely bonded systems\cite{natchem} (including the intermediate-range London dispersion interaction) around equilibrium, 
being comparable to or even more accurate than a computationally more expensive hybrid GGA (defined below).\cite{scan_solid}
However, SCAN is still 
a semilocal functional which inevitably fails for systems where the long-range effects are important, such as in
the self-interaction error encountered in stretched $H_2^+$ and long-range van der Waals interactions. 

Mixing part of the semilocal exchange with nonlocal Fock exchange
can reduce the self-interaction error, and is the dominant approach in quantum chemistry.
These hybrid DFAs were originally introduced by Becke 
and are motivated by the adiabatic connection.\cite{bhlyp}
Similarly, double-hybrid DFAs use the virtual orbital space to construct an approximate 
correlation energy.\cite{b2plyp,pwrb95}
Hybrid and double-hybrid DFAs are more computationally demanding than semilocal functionals.
While \mbox{(meta-)}GGAs scale as $N^3$, where $N$ is the size of the orbital basis, hybrids and double-hybrids
scale as $N^4$ and $N^5$, respectively.
Hybrid and double hybrid variants of SCAN have also recently been reported.\cite{scan0}

The long-range London dispersion interactions (also known as attractive van der Waals forces) 
are important for describing extended systems such as
condensed hard and soft matter, larger molecular assemblies, or adsorption processes on various surfaces.
For reviews or overviews on the ``dispersion problem in DFT'', see Refs.~\onlinecite{burns11,klimes12,dftd_wire}.
In this study we show how to combine the SCAN meta-GGA with modern London dispersion corrections.
While we will focus on the most efficient D3 scheme by Grimme and coworkers\cite{dftd3}, 
we will also consider the VV10 nonlocal density kernel by Vydrov and Van Voorhis.\cite{vv10}
A related SCAN+rVV10 scheme, where rVV10 stands for a revised VV10\cite{rvv10}, has also been developed and yields
excellent accuracy for predicting properties of layered materials.\cite{scanvv10}

Due to their computational efficiency, the (meta-)GGA DFAs are heavily relied on for the computation of
geometries. For other properties (e.g., band gaps of solids), more accurate results from 
hybrid and double hybrid DFAs\cite{b2plyp,gmtkn30}
or even high level (local) coupled cluster methods are needed.\cite{ao_lmp2,schuetz_lcc,dlpnoccsdt}
Specifically for condensed phases, geometry optimizations with a hybrid DFA 
using large orbital basis sets are not amenable for routine applications.
Typical organic crystals with $200$ atoms in the unit cell need about $10^5$ plane wave basis functions.\cite{tbtq}
In this orbital space, the computation of Fock exchange is very demanding.
In systems with local electron density,
an atom-centered Gaussian basis set can be used to reduce the computational cost, 
which was one of the design strategies of the PBEh-3c composite method.\cite{pbeh3c}
However, basis set errors have to be compensated\cite{sure_smallbasis} 
and small gap systems are problematic.
Thus, a meta-GGA with improved equilibrium geometries is desired.

We begin with a short methodological description in section~\ref{sec:method}.
Consequences of the sensitivity with respect to integration grids sometimes encountered for meta-GGAs\cite{jwd04,jbs09} 
are discussed in section~\ref{subsec:scan}.
Then, the D3 and VV10 London dispersion corrections are described 
and the recommended damping parameters are given (section~\ref{subsec:disp}).
In section~\ref{sec:results} we examine the accuracy of the combined SCAN-D3 method and give a broad overview
on various covalent and noncovalent bonding regimes (section~\ref{subsec:geom}).
In addition, noncovalent interaction energies and some main group thermochemistry 
and kinetics are analyzed in sections~\ref{subsec:nci} and \ref{subsec:gmtkn}.

\section{\label{sec:method}Methodology}
\subsection{\label{subsec:scan}The SCAN meta-GGA}
A general meta-GGA form for the xc energy can be written as
\begin{align}
\label{eq:mgga-xc}
  E_{xc} = & \int \textrm{d}\mathbf{r}\, 
      f(\rho(\mathbf{r}),
        \gamma(\mathbf{r}),
        \tau(\mathbf{r})) \,,
\end{align}
where we define $\gamma(\mathbf{r}) = \nabla\rho(\mathbf{r}) \cdot \nabla\rho(\mathbf{r})$.
SCAN improves upon previous nonempirical meta-GGAs such as TPSS and MGGA-MS\cite{ms0} by satisfying more exact constraints 
on the xc energy and by resolving the ``order of limits'' problem\cite{ptss04} encountered for meta-GGA parametrizations of $f$
using both of the $\tau$-dependent variables $z=\tau^{\text{vW}}/\tau$ and $\alpha$, defined below, where
$\tau^{\text{vW}} = |\nabla \rho|^2/8\rho$ is the von Weizs\"{a}cker kinetic energy density.
Instead SCAN utilizes only the $\tau$-dependent variable $\alpha = (\tau - \tau^{\text{vW}})/\tau^{\text{unif}}$ to identify different density 
regimes such as those found in covalent ($\alpha = 0$), metallic ($\alpha \approx 1$), and weak ($\alpha >> 1$) bonds.
$\tau^{\text{unif}} = (3/10)(3\pi^2)^{2/3} n^{5/3}$ is the kinetic energy density of a uniform electron density.
By incorporating ``appropriate norms'', systems where the exact xc hole is localized near the reference electron, 
SCAN is pushing the limits of accuracy achievable by a nonempirical semilocal functional.

The underlying parametrization of SCAN can be summarized into three steps for the exchange part.  
Uniform density scaling is first satisfied by
using the reduced density gradient variables $s$ and $\alpha$ to parametrize $f$. An approximate form appropriate for 
$\alpha \approx 1$ was then developed such that it satisfies the fourth-order gradient expansion for exchange.
An approximation for $\alpha = 0$ was designed to recover
the exchange energy of the hydrogen atom in addition to satisfying several other limits.
The final form is then constructed as an interpolation to connect $\alpha=0$ and $\alpha \approx 1$ 
and an extrapolation to extend to $\alpha \rightarrow \infty$ while satisfying all possible exact constraints.
A similar procedure was used for correlation resulting in an interpolation and extrapolation that yields vanishing
correlation energy for any one-electron density along with several other relevant limits of the correlation energy for both
atoms (or molecules) and solids. A full list of the constraints satisfied by SCAN for both exchange and correlation
can be found in the Supporting Information of Ref.~\onlinecite{scan}.
While the choice of only $\alpha$ over $z$ and $\alpha$ is desirable for the reasons stated above, it leads to some complications
in the numerical integration of the SCAN potential.

Using Eq.~\eqref{eq:mgga-xc} we can express a matrix element 
for atom-centered basis functions $\chi_\mu$ and $\chi_\nu$
of the meta-GGA xc potential within the Neumann, Nobes, Handy approximation\cite{nnh96} as
\begin{align}
   V^{xc}_{\mu\nu} = & \frac{\partial E_{xc}[D]}{\partial D_{\mu\nu}} = 
     \int \textrm{d}\mathbf{r}\,  \left[
         \frac{\partial f}{\partial \rho}(\mathbf{r})
         \chi_\mu(\mathbf{r})
         \chi_\nu(\mathbf{r}) 
       \right. \\ \nonumber   & \left.
     + \left( 
         2\frac{\partial f}{\partial \gamma}(\mathbf{r})
         \nabla\rho(\mathbf{r})
       \right) \cdot \nabla(\chi_\mu(\mathbf{r})\chi_{\nu}(\mathbf{r}))
       \right. \\ \nonumber & \left.
     + \frac{\partial f}{\partial \tau}(\mathbf{r})
       \nabla\chi_\mu(\mathbf{r}) \cdot \nabla\chi_\nu(\mathbf{r})
     \right] \, ,
\end{align}
where $D_{\mu\nu}$ is the KS density matrix and partial-integration has been used to 
avoid second-derivatives of the basis functions.\cite{fp06}
Numerical challenges can arise when evaluating the term 
proportional to $\partial f/\partial \tau$.  Previous works have shown that meta-GGA potential
energy surfaces for dispersion bound complexes can exhibit spurious 
oscillations using too small grids\cite{jbs09}, and that reaction energies can be severely impacted 
by the choice of grid as well.\cite{wh10}
Terms proportional to this function also arise for molecular properties such as nuclear gradients,
see, e.g., Eq.~(15) of Ref.~\onlinecite{fp06},
and hence analytic geometry optimizations are also influenced by the choice of integration grid.
For atoms, the derivative of the SCAN energy density can exhibit oscillations near $\alpha \approx 1$ due to
its functional form\cite{mggaoep}, so we report the convergence behavior of the energy and nuclear gradient with respect to the 
numerical integration grid for completeness.
Since numerical grids used to evaluate DFT contributions are built by combining angular and
radial grids, we have studied the impact of convergence in both grids separately. 
For a given angular integration grid, slow convergence of 
the total energy and nuclear gradient with respect to the radial integration grid was encountered.
To accurately integrate the SCAN potential,
a significantly larger number of radial points are needed in \textsc{turbomole}
compared to previous nonempirical functionals such as TPSS.
Using a converged radial grid, however, the convergence of the angular grid is typically 
much faster and sufficiently accurate results can be obtained using grid 4 in \textsc{turbomole}
which is only slightly larger than the default (grid m3).
We report more detailed information on the grid dependence in the supporting information\cite{suppinfo},
the conclusions of our tests being that energy differences are less sensitive to the grid issue
than nuclear gradients. Therefore, in practice a very large radial grid is only required when computing 
molecular properties, and not necessarily for computation of typical reaction energies 
which can be adequately described using a slightly augmented radial grid.

\subsection{\label{subsec:disp}London dispersion interaction}

To obtain long-range corrections from DFT, we start with the adiabatic fluctuation dissipation theorem
to formulate an exact expression for the correlation energy
\small
\begin{align}
 \label{eq:acfdt}
E_{\text{c}} = -\frac{1}{2\pi} &\int_{0}^{1} \!\textrm{d} \lambda \int \!\textrm{d} \mathbf{r} \textrm{d} \mathbf{r}' \frac{1}{|\mathbf{r}-\mathbf{r}'|} \notag \\
\times
& \int_{0}^{\infty} \!\textrm{d} \omega \left[  \chi_{\lambda}(\mathbf{r}, \mathbf{r}', i\omega)
- \chi_{0}(\mathbf{r}, \mathbf{r}', i\omega) \right] \, ,
\end{align}
\normalsize
where the Coulomb interaction is scaled by $\lambda$ and $\chi_{\lambda}$ 
is the corresponding dynamical charge density susceptibility (also called the linear response function).\cite{acfdt_a,acfdt_b}
The linear response of the noninteracting KS system is fully described by $\chi_{0}$ 
with the occupied and virtual KS orbitals $\psi_i$ and  $\psi_a$, respectively,
\small
\begin{align}
\chi_{0}(\mathbf{r}, \mathbf{r}', i\omega) = -4 \sum_{i}^{} \sum_{a}^{} \frac{\omega_{ai}}{\omega^2_{ai}
+ \omega^2} \psi_i(\mathbf{r}) \psi_a(\mathbf{r}) \psi_a(\mathbf{r'}) \psi_i(\mathbf{r}') \, . 
\end{align}
\normalsize
Here, $\omega_{ai}$ denotes the orbital energy differences corresponding to an excitation from orbital $i$ into orbital $a$.

The computation of the interacting response function is a difficult problem and so
one has to derive certain approximations in order to obtain a tractable correlation energy, which is
done in all modern London dispersion corrections as outlined in two recent review articles.\cite{grimme_chemrev,tkatchenko_rmp}
One example is the VV10 nonlocal density kernel by Vydrov and Van Voorhis\cite{vv10,vv10_b,vv10_c,vv10_d}, where
the susceptibility is replaced by an approximation based on the local density $\rho$ and its reduced gradient $s$
\small
\begin{align}
 E^{\text{VV10}}_{c}  =& \int\! \mathrm{d} \mathbf{r} \rho(\mathbf{r})
\left[ \beta + \frac{1}{2} \int \!\mathrm{d}\mathbf{r}' \rho(\mathbf{r}') \Phi(\mathbf{r},\mathbf{r}') \right] \,,\notag\\
\Phi(\mathbf{r},\mathbf{r}') =& \Phi(\mathbf{r},\mathbf{r}',\rho(\mathbf{r}),\rho(\mathbf{r}'),s(\mathbf{r}),s(\mathbf{r}');C,b)\,.
\end{align}
\normalsize
Within this method the Coulomb operator is treated within the dipole approximation and the frequency dependence
is estimated via the local plasmon frequency.
Two parameters are needed to determine the model; the first ($C$) is adjusted to reproduce reference 
dispersion coefficients at large distances, and
the second ($b$) is used to damp the VV10 contribution at short distances.
The parameter $b$ can be used to adjust the VV10 kernel to any semilocal DFA.\cite{hujo2011_1}

To derive the working equations for the D3 scheme, further partitioning has to be done.
By spatially integrating the response function,
we can define the corresponding polarizability tensor $\alpha_{ij}(i\omega)$ of a fragment
\begin{align}
 \alpha_{ij}(i\omega) = \int\! \textrm{d} \mathbf{r} \textrm{d} \mathbf{r}'\, \mathbf{r}_i \mathbf{r}'_j \chi(\mathbf{r},
\mathbf{r}', i\omega) \, .
\end{align}
The most natural fragments in a molecule are the individual atoms since,
due to their spherical symmetry, only an isotropic dynamical polarizability has to be considered.
Thus, the correlation energy between two atoms (A and B) can be expressed by the Casimir-Polder relation\cite{casimir1948}
\begin{align}
\label{eq:casimir-polder}
  E^{\text{AB}}_{c} =& -\frac{C_6^{AB}}{R_{AB}^6}\,,\notag\\
 C_{6}^{AB} =& \frac{3}{\pi} \int\limits_{0}^{\infty}\!\textrm{d}\omega\, \alpha_{A}\left(i\omega\right) \alpha_{B}\left(i\omega\right)\,.
\end{align}
This is the leading order fluctuating-dipole--fluctuating-dipole term; 
similar higher-order contributions in both the multipole and the many-body sense can be constructed.
The most significant difference between the various dispersion correction schemes
is the way in which the $C_6$ coefficients are estimated.
In the Tkatchenko-Scheffler (TS)\cite{ts} method (and its extension MBD\cite{mbd}), the atomic $C_6$ coefficients 
are obtained via reference values from the free atoms
and scaled by the relative Hirshfeld volume of the atom in a molecule.
In the exchange-hole dipole moment (XDM) model, the exchange hole is integrated locally to yield atomic dipoles, 
which are then related to the $C_6$ coefficients.\cite{xdm_a,xdm_b,xdm_c,xdm_d}
For the D3 scheme, the dynamic polarizabilities of reference systems (hydrated atoms) are calculated via time-dependent DFT,
and a modified Casimir-Polder integration (similar to Eq.~\ref{eq:casimir-polder}) yields the 
atom-pair $C_6^{AB}$ value.\cite{dftd3}
A fractional coordination number is used to interpolate between the reference points.
Higher-order dipole-quadrupole pair-terms and dipole-dipole-dipole three-body terms (Axilrod-Teller-Muto type\cite{atm_a,atm_b})
are calculated via recursion relations and averages, respectively, from the corresponding $C_6$ coefficients.
The importance of many-body dispersion interactions has been recently analyzed by various groups.\cite{tkatchenko_mbdrev,dobson_mbd,sherrill_mbd}

In this work, the D3 scheme is always used including the three-body term. 
Together, the D3 contribution to the interaction energy is
\small
\begin{align}
 E_{\textrm{c}}^{(\textrm{D3})} =& -\frac{1}{2}\,\sum_{n=6,8}\sum_{A,B}^{\textrm{pairs}} \frac{C_n^{AB}}{r_{AB}^n} f_n^d(r_{AB})\notag\\
  &- \frac{1}{6}\,\sum_{A,B,C}^{\textrm{triples}}  \frac{C_9^{ABC}\left(1+3\cos\theta_A\cos\theta_B\cos\theta_C\right)}{r_{ABC}^9} f_9^{d}(r_{ABC})
\end{align}
\normalsize
The damping functions $f_n^d$ are introduced to combine the D3 dispersion interaction with the 
semilocal correlation contribution from the DFA. 
While the three-body term is damped to zero at short-range ($f_9^{d}(0)=0$) and kept fixed,
the two-body damping $(f_{6,8}^d)$ can be either used with a zero damping (one free parameter $rs_6$)
or a rational (Becke-Johnson) damping (two free parameter $a_1$ and $a_2$).\cite{dftd3bj}
Additionally, the dipole-quadrupole $C_8$ terms can be scaled by a parameter $s_8$.
Comparisons of the D3 with the VV10 dispersion correction
revealed very similar accuracies.\cite{vv10_vs_d3,vv10_bench}
While the VV10 scheme can adjust better to unusual electronic structures with strong charge transfer character,
the D3 dispersion coefficients are typically better for rather unpolar organic molecules.
Furthermore, the three-body term is available with highly efficient analytical derivatives,
which is important for large and dense systems.\cite{moellmann14}
\begin{table}[htb]
\caption{\label{tab:param}
Optimized damping parameter of the D3 and VV10 dispersion correction 
for the SCAN functional in comparison with other methods.}
\scalebox{0.85}{ 
\begin{ruledtabular}
\begin{tabular}{lrrrr}
          &  SCAN &  M06L & TPSS  & PBE0  \\\hline
\multicolumn{5}{c}{{\bf plain} (without correction)}\\    
$\delta E$\footnotemark[1] / \% & 22.7 & 14.4 & 56.5 & 43.6\\ 
$\delta R$\footnotemark[2] / \% & 1.2 & 0.5 & 14.5 & 7.3\\\hline
\multicolumn{5}{c}{{\bf D3} (default rational damping\cite{dftd3bj})}\\
    $s_8$ &  \footnotemark[3]0 &  --   & 1.9435 &  1.2177 \\
    $a_1$ &  \footnotemark[4]0.5380 &  --   & 0.4535 &  0.4145 \\
    $a_2$ &  5.4200 &  --   & 4.4752 &  4.8593 \\
$\delta E$\footnotemark[1] / \%
          & 7.7     & --    & 5.8    & 10.1\\ 
$\delta R$\footnotemark[2] / \%
          & 0.8     & --    & 1.7    &  1.1\\ \hline
\multicolumn{5}{c}{{\bf D3(0)} (zero-damping\cite{dftd3})}\\
    $s_8$ &  \footnotemark[3]0  &  \footnotemark[3]0 & 1.1050 &  0.9280\\    
    $rs_6$&  \footnotemark[4]1.3240 &  1.5810  & 1.1660 &  1.2870 \\
$\delta E$\footnotemark[1] / \% & 7.3 & 9.2 & 6.3 & 12.9\\ 
$\delta R$\footnotemark[2] / \% & 1.0 & 0.5 & 1.5 & 0.9\\ \hline
\multicolumn{5}{c}{{\bf VV10} (zero-type damping\cite{vv10})}\\    
    $b$   &   \footnotemark[4]14.0 &  18.9  & 5.0 &  6.0 \\
$\delta E$\footnotemark[1] / \% &  8.4 & 8.0 & 6.3 & 15.6\\ 
$\delta R$\footnotemark[2] / \% &  0.9 & 0.9 & 1.2 & 0.8\\
\end{tabular}
\end{ruledtabular}
}
\footnotetext[1]{Mean absolute rel. deviation of the S66x8 equilibrium energies.\cite{s66}}
\footnotetext[2]{Mean absolute rel. deviation of the S66x8 equilibrium distances.\cite{s66}}
\footnotetext[3]{Value not fitted.}
\footnotetext[4]{This work.}
\end{table}

We have trained the above damping functions using
the S66x8\cite{s66} benchmark set.
It consists of 66 small to medium sized molecular dimers
at 8 center of mass distances (equilibrium geometry, 5 elongated distances, and 2 shortened distances)
with coupled cluster singles, doubles and perturbative triples reference energies
at the estimated single-particle basis set limit, CCSD(T)/CBS(est.).
Recently, more rigorously converged references have been presented that are used throughout this work.\cite{s66_martin}
We interpolate the potential surfaces and extract the equilibrium minimum
to compare with equilibrium binding energies and equilibrium distances
at the CCSD(T) level.
We fit the damping parameter by minimizing the weighted absolute relative deviations from the reference 
($\delta E + 10\delta R$).

A summary of the optimized damping parameter for the D3 scheme in both damping variants and the VV10 scheme
is given in Table~\ref{tab:param}.
We give the relative absolute deviations from the S66x8 reference minima
and compare with the M06L and TPSS meta-GGAs and the PBE0 hybrid functional.
Because the SCAN functional can cover medium-range correlation
to a high degree (similarly to M06L), the dipole-quadrupole term is set to zero.
Typical deviations of the various methods are 5-10\% for the interaction energy
and 0.5-2\% for the center of mass distance.
SCAN-D3 yields a good compromise of 8\% and 1\% error for the energy and the 
distance, respectively.
For the intrinsically very attractive Minnesota functionals,
the parameter fit of the rational damping function is not stable.
This double counting problem associated with the different damping functions
was recently investigated in detail.\cite{minnesota_disp}
SCAN also covers a large amount of medium range correlation,
but both damping variants can be successfully applied with very similar accuracy.

While we recommend the rational damping for TPSS-D3 and PBE0-D3, 
the Minnesota functionals have to be used with zero-damping.
We tested SCAN with both damping variants (for all benchmarks shown below) and obtained very similar results,
therefore we give only the results for the recommended rational damping scheme.
We will occasionally compare the results with the second-order 
M{\o}ller-Plesset perturbation theory (MP2)\cite{moller1934}, the simplest correlated wavefunction based method.
The performance of MP2 based methods can be substantially improved when the 
long-range dispersion contribution is replaced by a more accurate treatment. For instance, an
attenuated MP2 energy using the VV10 kernel for the dispersion part\cite{mp2v} has been developed,
as well as a method that replaces the dispersion contribution by coupling inclusive 
terms from time-dependent DFT\cite{mp2c,mp2c_pbc}
(which is closely related to the D3 dispersion coefficients, Eq.~\ref{eq:casimir-polder}).
If not stated otherwise, the defaults mentioned here are used throughout this study.

\section{\label{sec:results}Results}
\subsection{\label{subsec:geom}Geometries}

The analysis of molecular and condensed phase geometries is separated in the following way and closely follows the strategy in Refs.~\onlinecite{pbeh3c} and \onlinecite{pbeh3c_ref}:
first the covalent bond distances of different element classes are investigated (subsection~\ref{subsubsec:bonds}),
then we highlight the interplay between covalent bond distances and medium-range correlation in medium sized molecules (subsection~\ref{subsubsec:rot}),
and finally analyze the noncovalent binding distances of molecular dimers and solids (subsection~\ref{subsubsec:nci_dist}).

\subsubsection{\label{subsubsec:bonds}Bond distances}
Though the covalent bonds are mainly determined by the semilocal xc contributions from the DFA, 
we use the full methods (with London dispersion interaction), as the correction scheme should not deteriorate the covalent bonds.
In order to put the results into some broader perspective, we compare with results from M06L,\cite{m06l} TPSS-D3,\cite{tpss} and PBE0-D3.\cite{pbe0}
The M06L meta-GGA is used as the most prominent Minnesota DFA and applied without further correction as recommend by Truhlar and coworkers.\cite{peverati2014}
In the past years, Grimme and coworkers established the TPSS-D3 meta-GGA for computing most reliable geometries at rather low computational cost.\cite{tmc32,rot25}
Recently, extremely accurate geometries computed with the dispersion corrected hybrid functional PBE0-D3 have been reported.\cite{pbeh3c}
Due to the nonlocal Fock exchange, the hybrid PBE0-D3 has significantly higher computational costs compared to the other meta-GGA based methods.
We plot both the plain DFA (in gray) and the dispersion corrected variants (in color) to highlight the influence of the long-range correction for non-Minnesota functionals.

\begin{table}[htb]
\caption{\label{tab:bond}
Comparison of experimental and calculated
ground state equilibrium bond distances $R_e$ (in pm)
for 35 small first and second row molecules, third-row or higher main group molecules, and 3d-transition metal
complexes. (1~pm = 0.01~\AA)}
\begin{ruledtabular}
\begin{tabular}{lrrrrr}
measure&   SCAN-D3 &  SCAN & M06L & TPSS-D3 & PBE0-D3  \\\hline
\multicolumn{6}{c}{LMGB35 (first and second row molecules)\footnotemark[1]}\\
      MD\footnotemark[2] & -0.2  & -0.1  & -0.2  &  0.7 & -0.6\\
     MAD\footnotemark[3] &  0.6  &  0.5  &  0.5  &  0.8 &  0.9\\
      SD\footnotemark[4] &  1.1  &  0.8  &  0.8  &  0.6 &  1.3\\
     MAX\footnotemark[5] &  4.1  &  2.6  &  2.7  &  3.0 &  4.5\\[2ex]
\multicolumn{6}{c}{HMGB11 (third-row or higher main group molecules)\footnotemark[1]}\\
     MD                  &  0.1  &  0.2  &  2.6  &  1.7 & -0.3\\
     MAD                 &  1.0  &  1.0  &  3.6  &  1.8 &  1.0\\
      SD                 &  1.2  &  1.2  &  3.9  &  1.3 &  1.2\\
     MAX                 &  1.7  &  1.7  &  9.4  &  3.6 &  2.2\\[2ex]

\multicolumn{6}{c}{TMC32 (3d-transition metal complexes)\footnotemark[1]}\\
     MD                  & -1.8  & -0.9  &  0.3  & -0.3 & -0.5 \\
     MAD                 &  2.2  &  2.9  &  2.5  &  1.8 &  1.4 \\
      SD                 &  1.9  &  4.6  &  4.7  &  3.0 &  1.9 \\
     MAX                 &  7.2  & 19.2  & 19.9  & 12.6 &  7.0 \\
\end{tabular}
\end{ruledtabular}
\footnotetext[1]{See Ref.\cite{pbeh3c} for details.}
\footnotetext[2]{Mean deviation, $>0$ denotes too long bonds.}
\footnotetext[3]{Mean absolute deviation.}
\footnotetext[4]{Standard deviation.}
\footnotetext[5]{Maximum absolute deviation.}
\end{table}
\begin{figure}[hbt]
\includegraphics[width=0.49\textwidth]{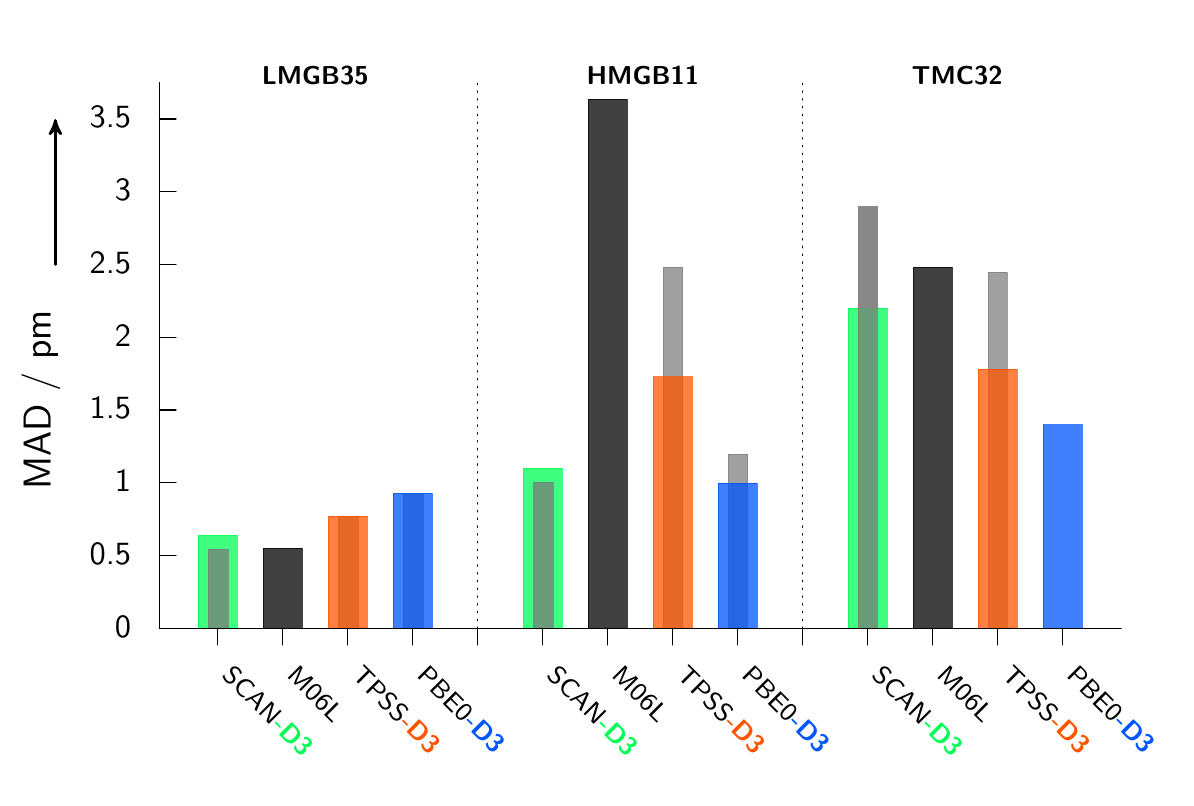}
\caption{\label{fig:bond} 
Mean absolute deviations of various methods for different bond distances
separated into light main group bonds (LMGB35), heavy main group bonds (HMGB11), 
and transition metal complexes (TMC32).
}
\end{figure}
In Table~\ref{tab:bond}, we report the comparison of experimental and calculated
ground state equilibrium bond distances $R_e$ (in pm) for 35 light main group bonds (LMGB35), 11 heavy main group bonds (HMGB11), and 32 3d-transition metal
complexes (TMC32).
The light main group bonds are sufficiently accurate with all applied methods, with 
mean absolute deviations (MADs) between 0.5 and 1.0\,pm.
SCAN-D3 provides solid results with an MAD of 0.6\,pm.
As expected, the dispersion correction has only a minor influence on these bond distances.
One main source of error is the partial multi-reference character of the F$_2$ and $F_2^{+}$ molecules, 
which are clear outliers for the hybrid functional PBE0-D3 (the MAD decreases to 0.7\,pm upon exclusion of the two systems).
The meta-GGAs implicitly account for some static correlation effects and hence are only slightly affected by these outliers.
TPSS-D3 is the only method with a systematic shift towards too long bonds, which is typical for most (meta-)GGA functionals.\cite{pbeh3c,rot25}
Compared to the plain Hartree-Fock (HF) mean field method, which has an MAD of 2.8\,pm, all semilocal DFAs lead to a substantial improvement.

The differences for the heavy main group bonds in the HMGB11 set are more pronounced.
The base line for a good method can be again defined by the HF MAD of 2.2\,pm.
While SCAN-D3 and PBE0-D3 provide excellent results with MADs slightly below 1.0\,pm, 
the error increases to 1.9 and 3.6\,pm for TPSS-D3 and M06L, respectively. M06L has some strong outliers leading to a larger spread of errors, about 4\,pm.
The largest deviation of M06L occurs for Pb$_2$Me$_6$, where the bond length is overestimated by more than 9\,pm.

The TMC32 set of 3d-transition metal complexes is particularly interesting as its description with hybrid functionals is rather problematic.\cite{tmc32}
This can also be seen by the bad performance of HF with MAD larger than 12\,pm.
In contrast, meta-GGAs are the ideal choice as they do not suffer from the inclusion of HF exchange for (organo-)metallic systems and
implicitly account for static correlation effects.
While TPSS-D3 performs excellently, the 2.2\,pm MAD of SCAN-D3 is very reasonable, outperforming both M06L and the uncorrected SCAN.
Due to the larger systems, the impact of the dispersion interaction is significant.
The error spread of SCAN-D3 is as small as the best performing PBE0-D3, but 
the bonds are systematically too short. The standard deviation drops by a factor of 2.4 when including the dispersion correction.
This indicates that the D3 scheme not only leads to a systematic shift (more strongly bound systems with shorter bonds),
but rather to an overall systematic improvement.

Concerning the bond lengths, the new SCAN-D3 functional provides very promising results.
It clearly outperforms the TPSS-D3 functional for all main group bonds and is of similar quality for transition metal complexes.
Compared to the popular M06L the bond lengths seem to be more reliable especially for heavier elements as seen in the HMGB11 benchmark set.

\subsubsection{\label{subsubsec:rot}Rotational constants}

In order to account for zero-point vibrational effects in the determination of molecular structures,
gas phase rotational spectra can be measured very accurately at low temperature.
From these spectra, the rotational constants, corresponding to inverse moments of inertia of the molecule,
can be extracted and used to infer structural information.
The accuracy of these measurements makes them an ideal benchmark observable to compare with
high-level quantum-chemical calculations\cite{rot_ccsdt}
and density functional approximations.\cite{rot25}
The rotational constants of small molecules can be calculated with an MAD of only 0.04\,\% 
using coupled-cluster methods accounting for up to quadruple excitations 
in conjunction with cc-pV6Z basis sets and corrections for core correlation.
A more cost-efficient CCSD(T)/cc-pVTZ calculation has a significantly higher error with MAD of 0.8\,\%.\cite{rot_ccsdt}
For molecules with more than a few heavy atoms, this is still a tremendous computational effort and 
it is important to have more efficient methods, such as the Random Phase Approximation\cite{rpa_geom}
that can robustly generate high-quality geometries.

A recently published set of 12 medium sized molecules has been corrected 
for anharmonic zero-point effects and can be directly compared to free optimizations.\cite{rot34}
\begin{table}[htb]
\caption{\label{tab:rot}
Comparison of experimental and calculated
geometries of medium sized molecules as judged by deviations from the rotational constants of ROT34.}
\begin{ruledtabular}
\begin{tabular}{lrrrrr}
measure                 &   SCAN-D3 &  SCAN & M06L & TPSS-D3 & PBE0-D3  \\\hline 
\multicolumn{6}{c}{ROT34 (rotational constants, deviations from \%)\footnotemark[1]}\\
      MD\footnotemark[2] &  0.12    &  0.06  &  0.15  &  1.18 &  -0.09\\
     MAD\footnotemark[3] &  0.24    &  0.27  &  0.28  &  1.18 &   0.27\\
      SD\footnotemark[4] &  0.29    &  0.36  &  0.36  &  0.50 &   0.33\\
     MAX\footnotemark[5] &  1.11    &  1.20  &  1.35  &  3.18 &   0.87\\
\end{tabular}
\end{ruledtabular}
\footnotetext[1]{See Ref.~\onlinecite{rot25,rot34} for details.}
\footnotetext[2]{Mean deviation, $>0$ denotes too large molecules.}
\footnotetext[3]{Mean absolute deviation.}
\footnotetext[4]{Standard deviation.}
\footnotetext[5]{Maximum absolute deviation.}
\end{table}
\begin{figure}[hbt]
\includegraphics[width=0.45\textwidth]{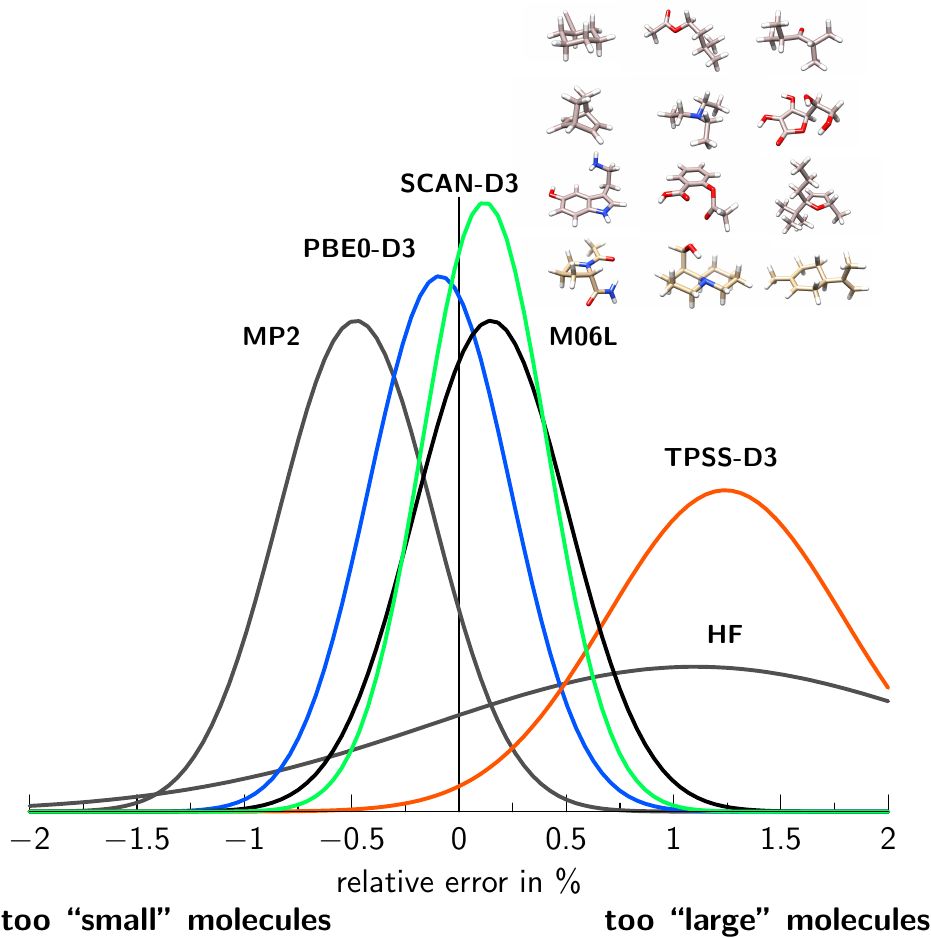}
\caption{\label{fig:rot} 
Normal distribution of the relative errors in the computed rotational constants $B_e$
for the ROT34 benchmark set with various theoretical methods. HF and MP2 results are taken from Ref.~\onlinecite{rot25}.
The inset shows the molecules of this set.
}
\end{figure}
Typical non-hybrid DFAs in the original study have MADs for the rotational constants 
of about 20--30\,MHz corresponding to 1--2\,\%.
MADs below 0.5\,\% could only be achieved with very few methods, 
all incorporating virtual excitations such as (spin component scaled) MP2
or the double hybrid functional B2PLYP.\cite{rot25} 
Later studies showed that a large amount of (effective) Fock exchange as in various long-range corrected
range-separated hybrid functionals can lead to similarly accurate results.\cite{rot34}
In a recent study, the excellent performance of PBE0-D3 was highlighted, which exceeds the accuracy of MP2.\cite{pbeh3c}

In Table~\ref{tab:rot}, the deviations of rotational constants with the reference computed with SCAN-D3
are shown alongside comparable methods.
For a visual comparison, the statistics are converted into normal error distributions in Fig.~\ref{fig:rot}.
The accuracy of both SCAN and M06L meta-GGAs is excellent
and exceeds the accuracy of all other tested (meta-)GGAs thus far. 
Since SCAN already covers medium range correlation to a high degree,
the impact of the dispersion correction is smaller, but still noticeable, compared to the more repulsive TPSS or PBE0.
Regardless, SCAN-D3 is the best performing method because
it has both a small MAD and standard deviation below 0.3\,\%.
Apparently it is possible to compute highly accurate molecular geometries using neither 
the virtual excitation space (dynamic correlation) nor the occupied orbital space in a nonlocal sense (Fock exchange).

\subsubsection{\label{subsubsec:nci_dist}Noncovalent distances}
In order to validate the accuracy of noncovalent distances, we use the above introduced S66x8 benchmark set of molecular dimers.\cite{s66}
The potential energy surface with respect to the center of mass distance is used to extract the equilibrium distance.
This is an ideal test since the molecular geometry is fixed at the reference MP2 level, so only the intermolecular interactions should influence the binding distance.
This has been recognized recently by several groups.\cite{headgordan_geom, pbeh3c,sure_smallbasis}

\begin{table}[h]
\caption{\label{tab:ncigeo}
Deviations of intramolecular center-of-mass distances $R_{CMA}$ from 
the CCSD(T) references for the S66x8 NCI benchmark set.
}
\begin{ruledtabular}
\begin{tabular}{lrrrrr}
measure&   SCAN-D3 &  SCAN & M06L & TPSS-D3 & PBE0-D3  \\\hline
\multicolumn{6}{c}{ S66x8 (CMA distance in \%)\footnotemark[1]}\\
      MD\footnotemark[2] &  0.0    &  0.2   & -0.4  &  1.4 &   0.3\\
     MAD\footnotemark[3] &  0.9    &  1.1   &  0.5  &  1.7 &   1.0\\  
      SD\footnotemark[4] &  1.2    &  1.4   &  0.8  &  1.9 &   1.3\\
     MAX\footnotemark[5] &  2.9    &  3.1   &  3.2  &  5.3 &   3.4\\
\end{tabular}
\end{ruledtabular}
\footnotetext[1]{See Ref.\cite{s66} for details.}
\footnotetext[2]{Mean deviation, $>0$ denotes too large distances.}
\footnotetext[3]{Mean absolute deviation.}
\footnotetext[4]{Standard deviation.}
\footnotetext[5]{Maximum absolute deviation.}
\end{table}
Repulsive density functionals like TPSS, PBE0, or plain HF fail to describe the dispersion bound systems.
The D3 dispersion correction reduces the MAD to 1.7\% and 1.0\% for TPSS-D3 and PBE0-D3, respectively.
This has to be compared to the accuracy of the computationally more expensive MP2 with MAD of about 1\%.
The main MP2 error originates from the bad description of $\pi$ systems, e.g., the equilibrium distance of the benzene
dimer is too short by 3.4\%.

The meta-GGAs M06L and SCAN-D3 perform excellently with MADs below 1\%. 
The dispersion correction slightly improves the behavior of SCAN, 
while the M06L performance would be deteriorated by an additional attractive contribution. 
M06L yields highly accurate geometries, but the corresponding systematic shifts are small 
and already indicate an overbound system with too dense molecular structures and too short noncovalent distances.
In contrast, SCAN yields systematically too large molecular structures and noncovalent distances as expected due to its lack of the long-range London dispersion interaction.
SCAN can therefore gain from the addition of a long-range D3 dispersion correction.
Indeed, the dispersion correction not only removes the systematic shift, 
but also reduces the error spread (apart from the LMBG) 
on all analyzed sets, indicating a physically sound contribution.

\subsection{\label{subsec:nci}Noncovalent interactions}
The above section demonstrates the excellent performance of SCAN-D3 for geometries. 
Its performance on the binding energies of several noncovalently bonded dimers and solids will now be discussed.

\subsubsection{Molecular dimers}
We use three standard benchmark sets introduced by Pavel Hobza and coworkers.
The first is the very well-known and widely used S22 set\cite{s22}
comprising 22 medium-sized molecules, mostly organic complexes in their equilibrium structure.
This set covers hydrogen bonded as well as typical vdW complexes and it has become the 
de-facto standard in the field of theoretical non-covalent interaction calculations. 
Note that the S22 reference values have been revised twice\cite{takatani2010,marshall2011} and we use the latest published values here.
The second is the already mentioned S66x8\cite{s66}, which is similar to the S22 with slightly larger complexes, 
less focus on hydrogen bonds, and accounts for some non-equilibrium structures.
Significantly larger complexes are compiled in the L7 test set\cite{l7}, and
we use the more consistent DLPNO-CCSD(T)/CBS$^{*}$ interaction energies as a reference.\cite{kruse2015} 
\begin{table}[hbt]
\caption{\label{tab:nci}
Deviations of intramolecular interaction energies from
the CCSD(T) references for the S22, S66x8, and L7 NCI benchmark sets. (1~kcal/mol = 0.0434~eV)}
\begin{ruledtabular}
\begin{tabular}{lrrrrr}
measure&   SCAN-D3 &  SCAN & M06L & TPSS-D3 & PBE0-D3  \\\hline
\multicolumn{6}{c}{ S22 (binding energy in kcal/mol)\footnotemark[1]}\\
      MD\footnotemark[2] & -0.4 &  0.6 &  0.8 &  0.0 & -0.3\\
     MAD\footnotemark[3] &  0.4 &  0.9 &  0.8 &  0.4 &  0.5\\
      SD\footnotemark[4] &  0.7 &  1.1 &  0.5 &  0.6 &  0.7\\
     MAX\footnotemark[5] &  2.6 &  2.9 &  1.7 &  1.5 &  1.8\\[2ex]
\multicolumn{6}{c}{ S66x8 (equilibrium binding energy in kcal/mol)\footnotemark[6]}\\
      MD                & -0.4  &  0.5 &  0.6 &  0.0 & -0.5\\ 
     MAD                &  0.5  &  0.9 &  0.6 &  0.3 &  0.6\\
      SD                &  0.6  &  0.9 &  0.7 &  0.4 &  0.6\\
     MAX                &  3.0  &  2.6 &  1.4 &  1.3 &  2.2\\[2ex]    
     \multicolumn{6}{c}{ L7 (binding energy in kcal/mol)\footnotemark[7]}\\
      MD                &  1.2  &  7.9 &  3.0 &  0.9 &  \footnotemark[8]1.4\\
     MAD                &  2.5  &  7.9 &  3.0 &  1.1 &  1.6\\
      SD                &  3.0  &  5.3 &  2.4 &  1.2 &  1.2\\
     MAX                &  4.7  & 15.6 &  6.3 &  2.8 &  3.0\\          
\end{tabular}
\end{ruledtabular}
\footnotetext[1]{See Ref.~\onlinecite{s22} for details.}
\footnotetext[2]{Mean deviation, $>0$ denotes underbound systems.}
\footnotetext[3]{Mean absolute deviation.}
\footnotetext[4]{Standard deviation.}
\footnotetext[5]{Maximum absolute deviation.}
\footnotetext[6]{See Ref.~\onlinecite{s66} for details.}
\footnotetext[7]{See Refs.~\onlinecite{l7,l7new} for details.}
\footnotetext[8]{Values replaced by PW6B95-D3.\cite{l7}}
\end{table}
\begin{figure}[hbt]
\includegraphics[width=0.49\textwidth]{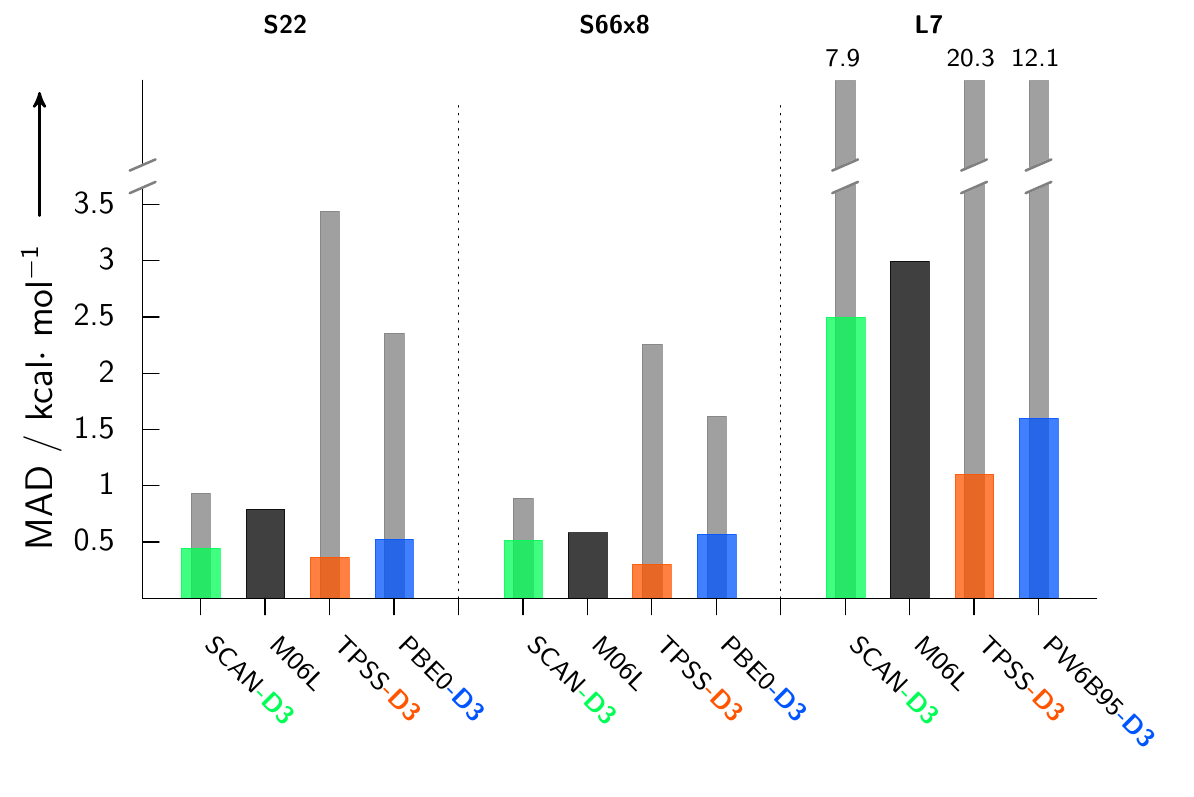}
\caption{\label{fig:nci} 
Mean absolute deviations of various methods for different molecular noncovalent interaction sets
separated into small dimers (S22), small to medium sized dimers at 8 center-of-mass distances (S66x8), 
and larger complexes (L7).
}
\end{figure}

The statistical deviations from the references are given in Table~\ref{tab:nci} and plotted in Figure~\ref{fig:nci}.
When comparing the different test sets, one has to keep in mind that the mean binding energies are 
7.3 kcal/mol,
5.5 kcal/mol, and
16.7 kcal/mol.
Nevertheless, we do not give relative deviations from the references as those would be dominated by only a few systems with tiny binding energy.

We confirm that TPSS-D3 is one of the most accurate non-hybrid DFAs for noncovalent binding energies of molecular complexes.
The MADs of 0.4, 0.3, and 1.1 kcal/mol for the S22, S66x8, and L7 sets are excellent and are all below 7\% of the mean binding energy.
The hybrid functional PBE0-D3 (and PW6B95-D3 for L7\cite{l7}) also performs very well for these sets.
The plain meta-GGA M06L has substantially larger errors and the MADs are approximately double compared to TPSS-D3.
While the errors could be reduced with the D3(0) scheme, this would simultaneously deteriorate the accurate geometries.
Consistent with the geometry analysis, SCAN is more repulsive compared to M06L and the performance even slightly worse.
This especially holds for the L7 test set, where the MAD is close to 50\% of the mean binding energy.
Clearly, the long-range part of the London dispersion interaction is missing and the results improve significantly
with the addition of the D3 correction. 
The results for the small and medium sized sets S22 and S66x8 are competitive with TPSS-D3.
The MAD on the L7 set with 2.5 kcal/mol (15\% of the mean binding) is reasonable,
but worse compared to some other methods.
It has been noted several times in the literature that, for highest accuracy on noncovalent energies between molecules,
the D3 and related semi-classical dispersion corrections have to be combined with intrinsically more repulsive DFAs.\cite{pernal2009,murray2009}
However, it is still notable that SCAN can profit from the dispersion correction and overall yields accurate noncovalent binding energies.

\subsubsection{Molecular crystals}
Molecular crystals are an increasingly important class of materials that require an accurate description
from efficient methods. 
This is especially important for 'in silico' crystal structure prediction.\cite{price2014,neumann2008,pantelides2014} 

To investigate this class of systems, we analyze the X23 set of (mostly) organic molecular crystals\cite{c21,x23} that
can be considered as a periodic extension of S66 where the asymptotic parts of the 
non-covalent interaction, specifically the dispersion component, may dominate.
The benchmark set X23 was compiled by Otero-de-la-Roza and Johnson\cite{c21} and further refined by Reilly and
Tkatchenko.\cite{x23} 
Experimental sublimation enthalpies are corrected for zero-point and thermal effects yielding 
electronic lattice energies which allow convenient benchmarking. The latter study also estimates the impact on anharmonic contributions to the 
sublimation energy and we use these values for benchmarking (see below).
One of the authors has back-corrected the X-ray unit cell volumes (or mass densities)
to yield zero-point exclusive observables.\cite{pbeh3c} 
These can then be directly compared to optimizations on the electronic energy surface.
In order to decrease the computational effort, we compiled a subset consisting of the crystals
cyclohexanedione,
acetic acid,
adamantane,
benzene,
CO$_2$,
cyanamide,
ethylcarbanate,
oxalic acid,
pyrazine,
pyrazole,
succinic acid, and
uracil.
The subset is constructed to maintain the MAD of TPSS-D3 for both the crystal density and the lattice energy within 0.5\%.

\begin{figure}[hbt]
\includegraphics[width=0.49\textwidth]{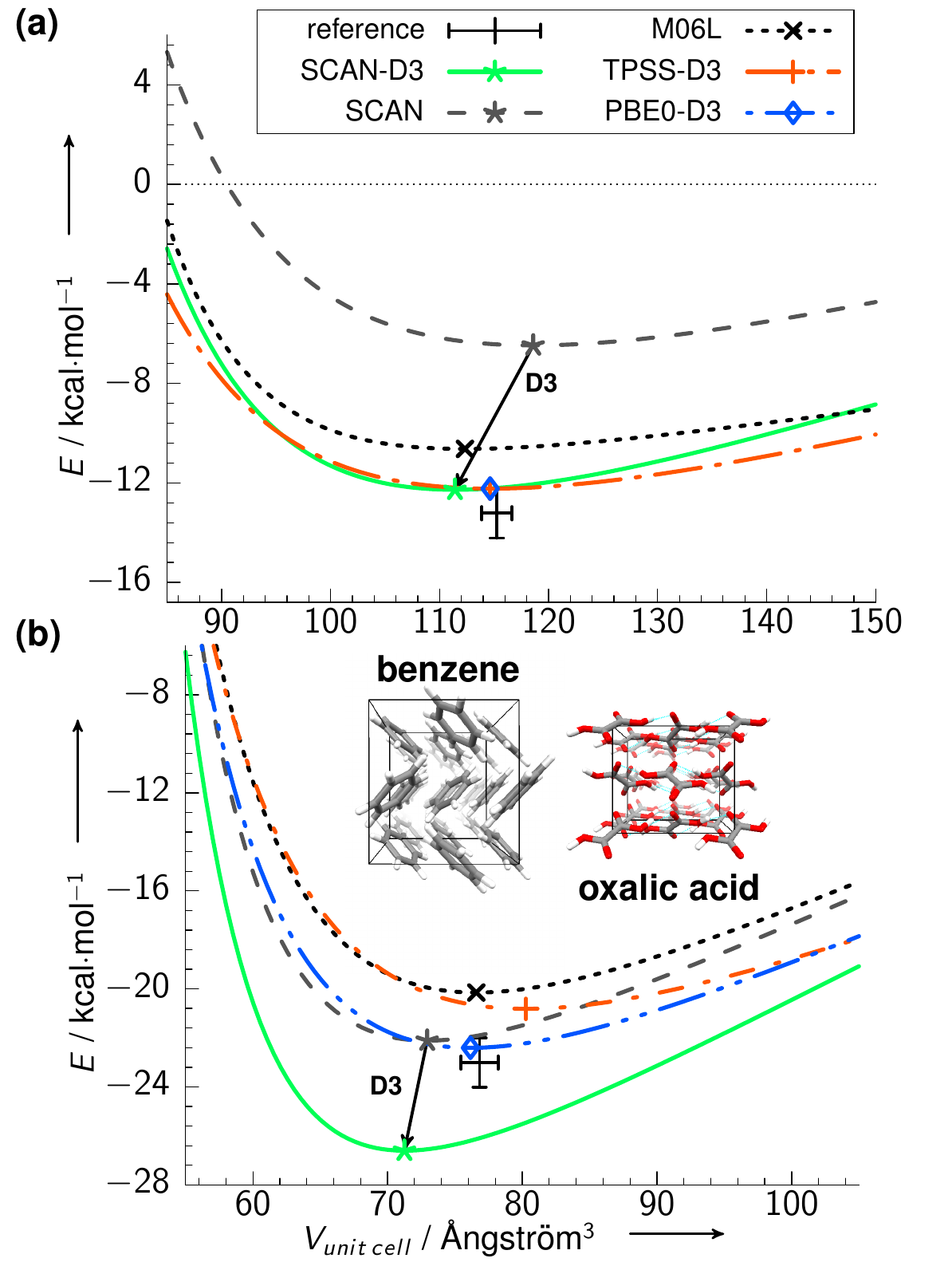}
\caption{\label{fig:pes} 
Lattice energy of the (a) benzene and (b) oxalic acid $\alpha$ crystal 
based on constrained volume optimizations (TPSS-D3 level) with
single-point evaluations of various dispersion corrected DFAs.
For each method, the cross shows the position of the energy minimum and the arrow indicates the effect of the added
dispersion correction.
}
\end{figure}

The statistical performance is summarized in Table~\ref{tab:molcryst}
and the potential energy surfaces of two selected crystals are shown in Figure~\ref{fig:pes}.
We show the PES of unpolar benzene and oxalic acid ($\alpha$ polymorph) that contain significant hydrogen bonds,
i.e, the contributions from electrostatic and induction effects
increase while the relative impact of London dispersion decreases.
\begin{table}[h]
\caption{\label{tab:molcryst}
Deviations of unit cell volumes and interaction energies from the back-corrected exp. reference for the X23 organic crystal set.}
\begin{ruledtabular}
\begin{tabular}{lrrrr}
measure&   SCAN-D3 &  SCAN & M06L & TPSS-D3 \\\hline
\multicolumn{5}{c}{X23 (unit cell volume in \%)\footnotemark[1]}\\
      MD\footnotemark[2] & -4.2    & -0.5   & -3.7  &  1.0 \\
     MAD\footnotemark[3] &  4.2    &  2.2   &  5.1  &  2.8 \\
      SD\footnotemark[4] &  1.6    &  2.5   &  4.1  &  4.0 \\
     MAX\footnotemark[5] &  6.6    &  4.7   &  8.4  & 15.0 \\[2ex]
\multicolumn{5}{c}{X23 (Lattice energy in kcal/mol)}\\
      MD &  1.5    & -3.7   & -1.2  & -0.7 \\
     MAD &  1.9    &  4.0   &  1.7  &  1.1 \\
      SD &  2.0    &  3.2   &  1.7  &  1.1 \\
     MAX &  5.0    & 10.5   &  3.4  &  2.2 \\
\end{tabular}
\end{ruledtabular}

\footnotetext[1]{See Ref.\cite{c21,x23,pbeh3c} for details.}
\footnotetext[2]{Mean deviation, $>0$ denotes too large distances.}
\footnotetext[3]{Mean absolute deviation.}
\footnotetext[4]{Standard deviation.}
\footnotetext[5]{Maximum absolute deviation.}
\end{table}
The PES of the benzene crystal in Figure~\ref{fig:pes}\,(a) shows again that both SCAN and M06L
already cover some part of the medium range dispersion interaction.
While the corresponding potentials show a clear minimum, the crystal is still underbound.
The minimum of SCAN-D3 is very close to the reference after adding the D3 correction and is
within ``chemical accuracy'' of 1 kcal/mol (similar to the TPSS-D3 and PBE0-D3 results). 
The equilibrium is a slightly more dense crystal than experimentally observed by about 3\%.

The oxalic acid crystal is one of the crystals within the X23 set with the strongest hydrogen bond contributions.
It is therefore much more challenging for a semilocal DFA to describe the induction effects accurately as shown in Figure~\ref{fig:pes}\,(b).
M06L is still underbound, but SCAN already computes a lattice energy close to the reference. 
Adding the D3 correction leads to a 4 kcal/mol overbinding with significantly too small unit cell volume.
Similarly to PBE, SCAN seems to overpolarize hydrogen bond networks leading to a too attractive induction interaction.
Adding the physically correct dispersion interaction enhances the overbinding tendency,
which leads to the comparably poor performance for the oxalic acid crystal.
A similar effect is seen for water clusters (WATER27,\cite{water27} see below), and ice polymorphs (ICE10\cite{ice10}).
Another study recently reported analogous behavior for SCAN on another set of ice polymporphs.\cite{scan_solid}
TPSS-D3 has smaller errors and is even bound too weakly and as expected the best results are computed with the hybrid PBE0-D3.

Benzene and oxalic acid are two borderline cases as the other X23 systems are typically in between them
as shown by the statistics given in Table~\ref{tab:molcryst}.
The SCAN-D3 MAD of 4.2\% for the unit cell volumes is worse compared to the uncorrected SCAN result
that we mainly attribute to the intrinsic errors of SCAN for hydrogen bonded systems. 
At the same time the standard deviation is slightly decreased 
indicating that though the D3 contribution is physically meaningful, the final SCAN-D3 method systematically underestimates the cell volumes.
The geometries at the M06L level are systematically too dense by about 3.7\% leading to an MAD larger than 5\%.
Clearly an additional dispersion correction would increase this systematic error.
TPSS-D3 yielded accurate unit cell volumes with an MAD below 3\%, the largest error occurring for the CO$_2$ crystal 
that is problematic for all dispersion corrected DFA methods.
The SCAN-D3 lattice energies have a reasonable MAD of 1.9 kcal/mol, the dispersion correction clearly improving the performance 
and lowering the MAD of SCAN by more than 50\%.
While the performance of M06L is similar, TPSS-D3 is significantly more accurate 
with an MAD close to 1 kcal/mol.
Other more repulsive DFAs (combined with various dispersion corrections) have been shown to yield analogous, highly accurate 
lattice energies on this X23 set, the most successful ones being PBE0-D3, PBE0-MBD, and B86PBE-XDM.\cite{c21,x23,moellmann14}

\subsection{\label{subsec:gmtkn}Thermochemistry and kinetics}
\begin{figure}[hbt]
\includegraphics[width=0.49\textwidth]{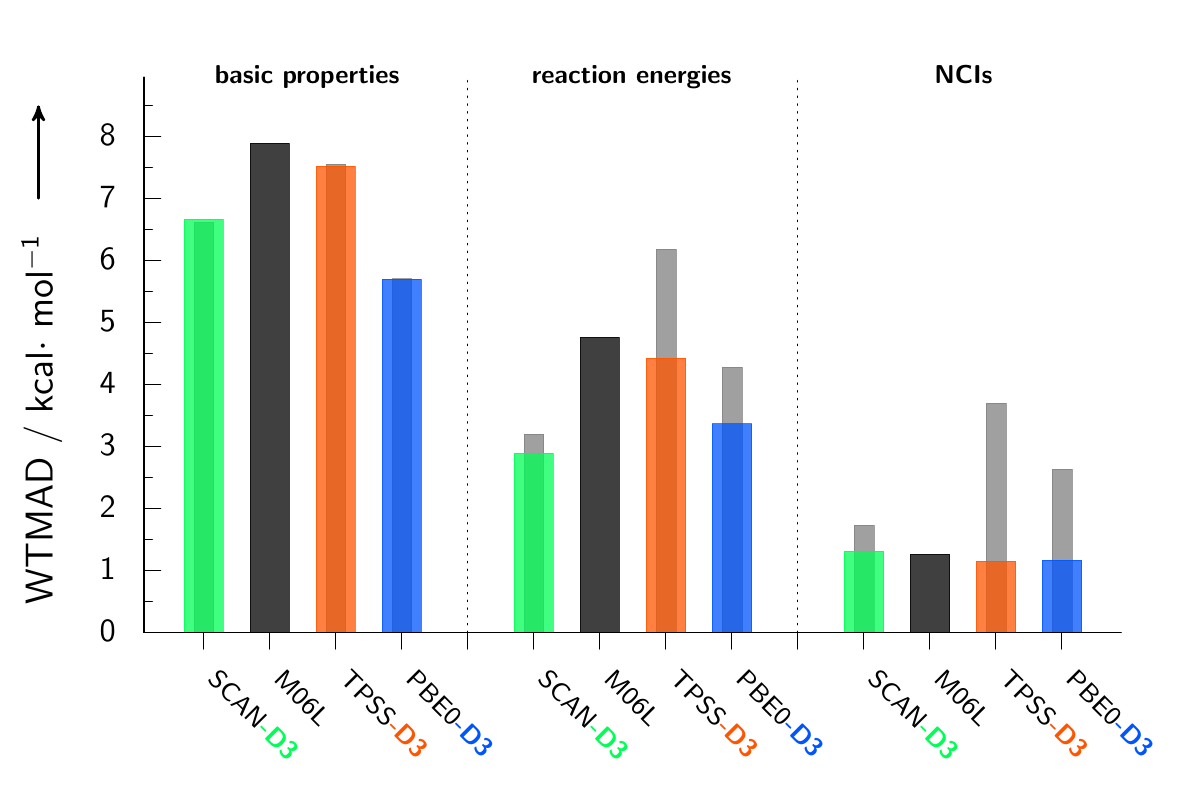}
\caption{\label{fig:gmtkn} 
Weighted mean absolute deviations (WTMADs) for the three categories (basic properties,
reaction energies, and noncovalent interactions) of the large GMTKN30 database composed of 30 individual benchmarks
sets given in Table~\ref{tab:gmtkn30}.
}
\end{figure}

In this final section we analyze the performance of the SCAN-D3 functional for
general main group chemistry.
In 2011, Goerigk and Grimme compiled a meta database of several benchmark sets, dubbed 
general main group thermochemistry, kinetics, and noncovalent interactions (GMTKN30).\cite{gmtkn24, gmtkn30}
It consists of three main subgroups testing basic properties 
(e.g., atomization energies, ionization potentials, electron and proton affinities, and reaction barriers),
reaction energies (including isomerizations),
and both intra and intermolecular
noncovalent interactions of light and heavy molecules, including molecular conformations.
This set has been extensively used to benchmark the large menagerie of DFAs from all different functional classes.\cite{gmtkn30_bench}
A transferable scheme to weight the different sets has been designed to compute an overall weighted mean absolute deviation (WTMAD),
enabling a direct comparison of all methods.
For all the tested functionals, the inclusion of a dispersion correction systematically improved the computed WTMAD.
Overall, the Jacob's ladder classification of DFA accuracy given in the introduction is borne out by these tests.
The inclusion of the virtual orbital excitation space is needed for highest routine accuracy (WTMAD below 2 kcal/mol).
Hybrid functionals employing the occupied orbital 
space with the nonlocal Fock operator lead to high accuracy (WTMAD below 3 kcal/mol) especially for 
reaction kinetics (e.g. energy barriers).
Typically semilocal meta-GGAs show only a small improvement compared to pure GGAs (WTMAD below 5 kcal/mol),
which are significantly better compared to LDA (WTMAD of 12 kcal/mol).
We compute the full GMTKN30 database with SCAN(-D3) and compare it to the meta-GGAs M06L and TPSS-D3
and the hybrid functional PBE0-D3 in Table~\ref{tab:gmtkn30}. The WTMADs of the three subgroups are shown in Figure~\ref{fig:gmtkn}.
Updated reference energies for the S22, WATER27, and HEAVY28 sets have been 
published\cite{marshall2011,water27_friedrich}, however we use the original references values of
Ref.~\onlinecite{gmtkn30_bench} so that the comparison of SCAN to previously published WTMADs is straightforward.

\begin{table}[hbt]
\caption{\label{tab:gmtkn30}
Mean absolute deviations (MAD, in kcal/mol) for all 30 subsets of the GMTKN30 database. 
Errors for M06L, TPSS-D3, PBE0-D3 are taken from Ref.~\onlinecite{gmtkn30_bench}.}
\begin{ruledtabular}
\begin{tabular}{lrrrrr} \hline
subset    &  \hspace{-1cm}SCAN-D3 &  SCAN & M06L & TPSS-D3 & PBE0-D3 \\\hline
\multicolumn{6}{c}{basic properties}\\
MB08-165  & 8.1 &  7.9 &13.3 & 9.5 & 8.6 \\       
W4-08     & 4.8 &  4.8 & 4.6 & 5.3 & 4.0 \\       
G21IP     & 4.9 &  4.9 & 4.5 & 4.0 & 3.7 \\       
G21EA     & 3.6 &  3.6 & 4.0 & 2.2 & 2.5 \\       
PA        & 3.2 &  3.2 & 4.6 & 4.7 & 2.8 \\       
SIE11     &10.2 & 10.0 &10.1 &11.6 & 7.8 \\       
BHPERI    & 3.8 &  3.2 & 3.5 & 3.1 & 1.6 \\       
BH76      & 7.9 &  7.8 & 3.8 & 9.0 & 4.4 \\[2ex]  

WTMAD (bp) & 6.7 &  6.6 & 7.9 & 7.5 & 5.7 \\[2ex]    

\multicolumn{6}{c}{reaction energies}\\
BH76RC    & 3.7 &  3.7 & 3.1 & 3.7 & 2.5 \\       
RSE43     & 1.9 &  1.9 & 3.1 & 2.2 & 1.8 \\       
O3ADD6    & 7.4 &  7.1 & 3.4 & 4.4 & 5.7 \\       
G2RC      & 6.8 &  6.6 & 5.9 & 6.8 & 6.8 \\       
AL2X      & 2.9 &  2.2 & 1.4 & 2.2 & 1.9 \\       
NBPRC     & 2.9 &  2.4 & 3.9 & 1.7 & 3.3 \\       
ISO34     & 1.3 &  1.4 & 2.2 & 2.1 & 1.6 \\       
ISOL22    & 4.2 &  4.6 & 7.4 & 7.0 & 2.9 \\       
DC9       & 8.6 &  8.8 &11.5 & 9.7 & 9.2 \\       
DARC      & 2.6 &  3.0 & 8.0 & 6.6 & 3.1 \\       
ALK6      & 3.8 &  3.4 & 8.1 & 3.3 & 3.6 \\       
BSR36     & 1.7 &  3.2 & 6.0 & 6.3 & 4.6 \\[2ex]  

WTMAD (re) & 2.9 &  3.2 & 4.8 & 4.4 & 3.4 \\[2ex]    

\multicolumn{6}{c}{non-covalent interactions}\\
IDISP     & 3.2 &  5.9 & 6.6 & 4.5 & 3.5 \\       
WATER27   & 9.4 &  7.4 & 2.8 & 4.9 & 6.4 \\       
S22       & 0.44 &  0.93 & 0.80 & 0.32 & 0.57 \\  
ADIM6     & 0.23 &  1.68 & 0.28 & 0.40 & 0.36 \\  
RG6       & 0.19 &  0.27 & 0.43 & 0.04 & 0.03 \\  
HEAVY28   & 0.28 &  0.40 & 0.65 & 0.20 & 0.17 \\  
PCONF     & 0.50 &  1.13 & 0.97 & 1.10 & 0.94 \\  
ACONF     & 0.16 &  0.32 & 0.46 & 0.05 & 0.10 \\  
SCONF     & 0.37 &  0.27 & 0.39 & 0.68 & 0.25 \\  
CYCONF    & 0.47 &  0.42 & 0.40 & 0.82 & 0.55 \\[2ex]  

WTMAD (nci)    & 1.3 &  1.7 & 1.3 & 1.2 & 1.2 \\[2ex]    
{\bf WTMAD }(all)    & {\bf 3.9} &  {\bf 4.1} & {\bf 4.9} & {\bf 4.6} & {\bf 3.6} \\
\end{tabular}          
\end{ruledtabular}
\end{table}
As expected, the effect of dispersion on the \emph{basic properties} of mostly small molecules is fairly small. 
The WTMAD of 6.7 kcal/mol is very good and in between the accuracy of a typical GGA and a hybrid functional.
Of this subgroup, the mindless benchmark (MB08-165) consisting of artificial molecules stands out.
The set has been designed to explore the breadth of chemical space and specifically 
analyze the DFAs far away from any training set to test their robustness.

The WTMAD of SCAN-D3 for the \emph{reaction energies} is excellent at 2.9 kcal/mol and surpases any other meta-GGA to date.
Even typical hybrid functionals like PBE0-D3 and B3LYP-D3 are worse with WTMADs of 3.4 and 4.7 kcal/mol, respectively.
The SCAN results for isomerization (ISO34, ISOL22) are particularly outstanding, and
in the subgroup of reaction energies the D3 dispersion correction leads to only small improvements.

Clearly, for the group of \emph{noncovalent interactions}, the dispersion correction has the largest impact by reducing 
the SCAN WTMAD from 1.7 kcal/mol to 1.3 kcal/mol. 
Compared to other functionals this reduction is moderate,
and especially intrinsically more repulsive DFAs can reduce the WTMAD below 1 kcal/mol (e.g. revPBE-D3).
The most problematic systems are the water clusters (WATER27), where the plain SCAN functional already 
leads to a too strong binding, which is then enhanced by the attractive dispersion contribution resulting
in the worst performance of the selected methods.
Hao {\it et al.} applied the meta-GGA made simple (MGGA-MS) with a D3 correction to the GMTKN30 set and found that 
it delivers top-notch performance for WATER27, with an MAD below 2~kcal/mol.\cite{gmtkn30_ms0} This illustrates that
it is possible to describe hydrogen bonds in water accurately via a nonempirical construction. In spite of the errors for 
WATER27, SCAN-D3 is still an overall improvement for non-covalent 
interactions compared to MGGA-MS, yielding a 1 kcal/mol smaller WTMAD for this subset.
Similar problems for water containing systems have been recognized for the PBE functional,\cite{ice10}
and are probably connected to an overpolarization problem in strong 
hydrogen bond networks related to the self-interaction error intrinsic to semilocal functionals.
On the other hand, SCAN-D3 is very accurate for molecular conformations 
with MADs below 0.5 kcal/mol for all 4 sets (PCONF, ACONF, SCONF, CYCONF).

\emph{Overall}, SCAN-D3 performs very well for the GMTKN30 with a WTMAD of 3.9 kcal/mol, one of the lowest for the meta-GGA class.
Interestingly, SCAN-D3 delivers superior performance compared to the M06L functional even
though parts of GMTKN30 are included in the training set of the Minnesota functionals.
Though MP2 can reduce the Hartree-Fock WTMAD of 18.3 kcal/mol to a satisfying 3.6 kcal/mol, the formal scaling
of MP2 ($N^5$) with respect to the single-particle basis set is much more demanding than that of SCAN-D3 ($N^3$).
Interestingly, a similar picture can be seen when comparing error statistics of molecular and atomic energies  with 
the method of atomic equivalents. The root mean square error on 592 species are
7.5, 4.7, and 4.2\,kcal/mol for LSDA, M06L, and SCAN, respectively,\cite{contraint_satisfaction}
reprocucing closely the trend shown by the GMTKN30 database.

\section{\label{sec:conlusion}Conclusions}
In this work we have combined the SCAN meta-GGA with a long-range correction for London dispersion interactions.
We provide default damping parameters for the D3(zero damping), D3(rational damping), and VV10 dispersion corrections.
The resulting SCAN-D3 method was tested on a broad set of systems with the main focus on accurate geometries,
as this represents the most advantageous aspect of the meta-GGA functional class.
Even considering hardware improvements, DFT will be the leading method to compute \emph{ab initio}
equilibrium structures in the foreseeable future.

The molecular geometries of SCAN-D3 exceed the accuracy of all other (meta-)GGAs thus far.
Rotational constants that measure the size of a molecule typically have a small MAD of 0.24\%, while
noncovalent binding energies are good (L7, X23) to very good (S22, S666x8),
producing high quality potential energy surfaces of molecular dimers and organic crystals.
Due to the self-interaction error intrinsic in semilocal functionals, SCAN, and thus SCAN-D3, 
overestimate hydrogen bonds in the same manner as PBE.
Thermochemistry and kinetics were shown to be in excellent agreement with reference values
as demonstrated on the large GMTKN30 database, resulting in a WTMAD of 3.9 kcal/mol.
Overall, SCAN-D3 delivers accurate properties that are close to the results of much more computationally demanding methods.
Importantly, this excellent performance has been achieved by 
a nonempirical semilocal functional. Indeed, all four of the functionals in Table~\ref{tab:gmtkn30} that
were not fitted to molecular data (except to a limited extent in the D3 correction) outperform
the one (M06L) that was, on the broad GMTKN30 database.

The long-range dispersion correction to SCAN is,
as expected, most important in systems that can have long-range dispersion binding, such as the
benzene crystal and the L7 set of large molecular complexes. As a consequence of the lack of
structure in the long-range correction, SCAN without D3 can be reasonably good for the 
geometry (but not the binding energy) of even the benzene crystal.

\small
\section{\label{sec:compdetail}Computational details}
For all molecular computations, we used a developer version of TURBOMOLE 7.0.\cite{turbomole_wire}
The M06L functional is computed via the XCfun interface.\cite{xcfun}
We use converged single-particle basis sets of quadruple-$\zeta$ quality (def2-QZVP).\cite{def2basen,qzvp}
Additional diffuse functions are used for the WATER27 and G21EA benchmark sets.\cite{aug-basis}
For heavy elements these are combined with the Stuttgart-Dresden effective core potentials,
that effectively include scalar relativistic effects.\cite{ecp2}
Only some hybrid PBE0 results that have been taken from previous work were evaluated with the slightly smaller def2-TZVP basis.
For the semi-local exchange-correlation part the numerical quadrature grids m4 (4 for SCAN) are used. 
For geometry optimizations with SCAN the radial grid size must be substantially 
increased to radsize 60 or 70, see the Supporting Information.
The RI-J approximation was used \cite{riold1,riold3,ridft} with default auxiliary basis sets.\cite{jauxbasnew}
Standard convergence threshold for SCF convergence ($10^{-7}$\,a.u.) and tight thresholds for geometry convergence ($10^{-4}$\,a.u.)
were applied.
Solid state calculations were conducted with a modified VASP5.3 program suite.\cite{vasp1,vasp2}
To approach the single-particle basis set limit, a projector-augmented 
plane-wave (PAW\cite{PAW1,PAW2}) basis set with a large energy cutoff of 1000\,eV was applied.
The PBE0 hybrid single-point energies (Figure~\ref{fig:pes}\,(b)) are calculated with a smaller energy cutoff of 500\,eV.
The Brillouin zone is sampled with dense $k$ grids of approximately $1/40$\,\AA$^{-1}$\ generated via the Monkhorst-Pack scheme.
For efficient geometry relaxations and three-body gradients of the D3 scheme in periodic boundary conditions,
we use a developer version of the CRYSTAL14 program.\cite{crystal14}

In the current Turbomole implementation, the requirements of SCAN's exchange-correlation functional on the numerical integration grid
are unusually high, leading  to an increased computational cost compared to TPSS by a factor of 2 to 10. However, SCAN has 
decreased numerical problems in VASP, where SCAN is only requires slightly denser Fourier grids compared to the PBE GGA.

\normalsize
We thank Stefan Grimme for helpful discussions concerning the D3 correction scheme 
and the importance of SCAN's numerical stability, and F. Furche for insight on the grid sensitivity in \textsc{turbomole}.
J.G.B. acknowledges support by the Humboldt foundation within the Feodor-Lynen program.
J.E.B. and A.R. were supported by the U.S. Department of Energy 
under grant \#DE-SC0010499. 
J.S. and J.P.P. were supported by the 
U.S. National Science Foundation under grant \#DMR-1305135 (with support from CTMC).

\bibliography{literature}

\end{document}